\documentclass[epj]{webofc}
\usepackage[varg]{txfonts}   
\wocname{EPJ Web of Conferences}
\woctitle{CONF12}
\usepackage[usenames]{color}
\usepackage{comment}
\usepackage{amsmath}

\def\beq{\begin{equation}}
\def\eeq{\end{equation}}
\def\bea{\begin{eqnarray}}
\def\eea{\end{eqnarray}}
\def\D0{D\O }

\begin{document}
\selectlanguage{english}
\title{Tensions in the flavour sector}
%
%

\author{Giulia Ricciardi\inst{1,2}\fnsep\thanks{\email{giulia.ricciardi@na.infn.it}}}

\institute{Dipartimento di Fisica E. Pancini, Universit\`a  di Napoli Federico II,
Complesso Universitario di Monte Sant'Angelo, Via Cintia,
I-80126 Napoli, Italy
\and
INFN, Sezione di Napoli,
Complesso Universitario di Monte Sant'Angelo, Via Cintia,
I-80126 Napoli, Italy
}

\abstract{%
 Flavour physics is currently well  described  by the Standard Model except for some measurements which could be signalling  new physics. We briefly summarize the status of tensions in rare exclusive semi-leptonic  $B$ decays and in $R_{K/D^{(\ast)}}$ ratios. We also address the issue of the long standing tension in the $|V_{cb}|$ and $|V_{ub}|$ exclusive/inclusive determinations.}
\maketitle
\section{Introduction}
\label{intro}
 In a general scenario of optimal agreement within the Standard Model (SM),   the flavour physics sector exhibits some measurements  which present non-significant but intriguing  tensions with SM predictions. Here we briefly discuss  $B$ decays mediated   by  the parton $b \to s \, l^+ l^-$ decays  at the lowest order in the SM. Being
 rare decays, they are  particularly susceptible to new physics effects. We also briefly address  the discrepancy between the $R_K$ and $R_{D^{(\ast)}}$ measured ratios and the corresponding SM predictions, which
 may hint at lepton non-universality. The last issue succinctly examined is  the determination of the  parameters of the Cabibbo-Kobayashi-Maskawa (CKM) matrix  $|V_{cb}|$ and $|V_{ub}|$, which
strongly affects the identification of new physics. The SM does not predict the values of the CKM matrix elements and their precise measurement allows a powerful check of the unitarity of the CKM matrix. A long standing tension remains between the  inclusive and exclusive determinations.

\section{Exclusive rare semileptonic $B$ decays}

LHCb has provided the most precise  measurements of the branching fractions of the $ B^+ \to K^{(\ast)+}   \mu^+ \mu^-$ and  $ B^0 \to K^{(\ast)0}   \mu^+ \mu^-$   decays \cite{Aaij:2014pli, Aaij:2012vr, Aaij:2013iag}.
All these measurements are below the
SM prediction.
%
The
 large $q^2$ region is the domain of election for
lattice QCD and
unquenched calculations of form factors  have been  performed. The more recent ones are   for $B \to K \, l l $ decays by HPQCD  \cite{Bouchard:2013pna, Bouchard:2013mia},    for $B_s \to K l \nu$ and $B \to K l l $ decays  by  Fermilab/MILC collaboration \cite{Liu:2013sya, Bailey:2015dka} and for $B_s \to \phi l^+ l^-$ and $B \to K^\star l^+ l^- $ decays by  the Cambridge  collaboration \cite{Horgan:2013hoa, Horgan:2013pva, Horgan:2015vla}.
Also in the light-cone-sum rule formalism, recent results have been presented, for $B \to \rho$, $B_q \to \omega$, $B\to K^\star$ and $B_s \to \phi$ form factors \cite{Straub:2015ica}.

Branching fraction measurements alone are
not sufficient to exploit the opportunities given by
exclusive rare semileptonic $B$ decays, which
present several
asymmetries and angular observables that can be studied as functions of the dimuon invariant mass squared, $q^2$.
Evidence of a non-vanishing value of
the isospin asymmetry \cite{Aaij:2012cq}
has not been confirmed in an update with larger statistics \cite{Aaij:2014pli}.
All
CP
asymmetries measured so far in these
decays are consistent with zero \cite{LHCb:2012kz, Aaij:2013dgw, Aaij:2014bsa}, as predicted
by the SM.

The angular distributions of the $ B^+ \to K^+  \mu^+ \mu^-  $ and  $ B^0 \to K^0_s  \mu^+ \mu^-$ decays
are described by two parameters, $F_H$, which is a measure of the contribution
from (pseudo)scalar and tensor amplitudes to the decay width  in the approximation that muons are massless, and
 $A_{FB}$, the
forward-backward asymmetry of the dimuon system.
 In the SM, ${\mathrm A_{FB}}$ is zero and ${\mathrm{F}_H} $ highly suppressed, and their  measured values are compatible with the SM expectations \cite{Aaij:2012vr, Aaij:2014tfa}.
 LHCb reports also the latest measurements of
 ${\mathrm A_{FB}} $ and $F_L$, the fraction of longitudinal polarisation, for the $B^0 \to K^{\ast 0} \mu^+ \mu^-$ channel
 as a function of the dimuon invariant mass, which agree with SM predictions \cite{Aaij:2011aa}.

A broad peaking structure has been observed in the dimuon spectrum of   $ B^+ \to K^+  \mu^+ \mu^-$ decays
 in
the kinematic region where the kaon has a low recoil against the dimuon system \cite{Aaij:2013pta}.  The mean and width of the resonance  are compatible with the
properties of the $\psi(4160)$.  The resonant decay and the interference
contribution make up 20\% of the yield for dimuon masses above 3770
MeV/c$^2$. This contribution
is larger than theoretical estimates \cite{Aaij:2013pta}.

In 2013,  new angular observables denoted as $P_{4,5,6,8}^\prime$,
that are free from form-factor uncertainties at leading
order \cite{Descotes-Genon:2013vna}, have been proposed in the $B^0 \to K^\ast \mu^+\mu^-$ channel. The LHCb experiment
has performed the first measurement  of these
angular observables using data collected in 2011 and announced  a 3.7$\sigma$
local discrepancy in one of the $q^2$ bins (4.30-8.68 GeV$^2$/c$^4$)
for  the angular observable $P_5^\prime$  \cite{Aaij:2013qta}.
This discrepancy has been confirmed in a new LHCb analysis  employing  the complete LHCb Run 1 dataset
 recorded in
$p p$
collisions at centre-of-mass energies of 7 and 8
TeV
during
2011 and 2012,  respectively, and  corresponding  to  an  integrated  luminosity  of  3.0 fb$^{-1}$
\cite{Aaij:2015oid}.
A deviation from the SM prediction in Ref. \cite{Descotes-Genon:2014uoa} has been observed in
each of the 4.0<$q^2$<6.0 GeV$^2$/c$^4$ and
6.0<$q^2$<8.0 GeV$^2$/c$^4$
bins at a level of 2.8 and
3.0 standard deviations, respectively. The LHC analysis \cite{Aaij:2015oid} also presents a complete set of observables, for the  first time, based
on the full angular distribution.   Correlations
between the different observables are computed to allow the results to be included in
global fits of $b \to s$ data. A global analysis of the
CP-averaged angular observables determined from the maximum
likelihood  fit indicates differences with the presently-available SM predictions at the level
of 3.4 standard deviations \cite{Aaij:2015oid}.
 These analyses have prompted a large number of theoretical investigations, searching for NP in  several frameworks or assessing  the effects of non-perturbative  corrections on the SM predictions, which may
change the significance of the discrepancy
(see e.g. \cite{Altmannshofer:2014rta, Beaujean:2013soa, Hurth:2013ssa, Mahmoudi:2016mgr, Hurth:2016fbr, Ciuchini:2016weo} and references within.)

\section{Exclusive decays into heavy leptons}

 Exclusive semi-tauonic $B$ decays were
first observed by the Belle Collaboration in 2007 \cite{Matyja:2007kt}.
Subsequent
analysis by Babar and Belle \cite{Aubert:2007dsa, Bozek:2010xy,Huschle:2015rga} measured
 branching fractions above, although consistent with, the SM predictions.
The ratio of branching fractions (the denominator is the average for $l \in \{e, \mu\}$)
\begin{equation}
R_{D^{(\ast)}} \equiv  \frac{{\cal{B}}( B \to D^{(\ast)} \tau \nu_\tau)}{{\cal{B}}( B \to D^{(\ast)} l  \nu_l)}
\label{ratiotau0}
\end{equation}
 is typically used instead of the absolute branching fraction
of $ B \to D^{(\ast)} \tau  \nu_\tau$ decays to cancel  uncertainties common to the numerator and the denominator.
These include the CKM matrix element $|V_{ub}|$ and several theoretical uncertainties on hadronic form factors and experimental reconstruction effects.
 In 2012-2013
Babar
 has measured
$R_{D^{(\ast)}}$ by using  its full data sample \cite{Lees:2012xj, Lees:2013uzd},
and reported a significant excess over the SM expectation, confirmed  in 2015
by LHCb \cite{Aaij:2015yra}.
In 2016 such excess has  been confirmed also by the Belle Collaboration, which has performed the first measurement of $R_{D^\ast}$
using the semileptonic tagging method, giving \cite{Abdesselam:2016cgx}
\begin{equation}
R_{D^\ast} = 0.302\pm 0.030\pm 0.011
\label{ratiotauBlle2016}
\end{equation}
where the first error is statistic and the second one is systematic.
%
By averaging the most recent measurements  \cite{Huschle:2015rga,Lees:2012xj, Lees:2013uzd,Aaij:2015yra, Abdesselam:2016cgx}, the HFAG Collaboration has found \cite{HFAG2016}
\begin{equation}
R_D  = 0.397 \pm 0.040 \pm 0.028  \qquad \qquad
R_{D^\ast}= 0.316 \pm 0.016 \pm 0.010
\label{ratiotau}
\end{equation}
where the first uncertainty is statistical and the second is
systematic. $R_D$ and $R_{D^\ast}$  exceed the SM
 value $R_{D}^{SM} = 0.300\pm 0.008 $ given by the HPQCD Collaboration \cite{Na:2015kha} and the SM phenomenological prediction $R_{D^\ast}^{SM}= 0.252\pm 0.003$ \cite{Fajfer:2012vx}
  by  1.9$\sigma$ and 3.3$\sigma$, respectively.
The combined analysis of  $R_D$ and  $R_{D^\ast}$, taking
into account measurement correlations, finds that the deviation
is 4$\sigma$ from the SM prediction.
Other recent SM predictions are available for $R_{D}^{SM}$,  that is $R_{D}^{SM} = 0.299 \pm 0.011$ by FNAL/MILC Collaboration \cite{Lattice:2015rga} and  $R_{D}^{SM} = 0.299 \pm 0.003 $ \cite{Bigi:2016mdz}. They are also below data, and in agreement with
older $ R_{D}^{SM}$ determinations \cite{Kamenik:2008tj, Becirevic:2012jf}.

Most recently, the Belle collaboration has reported
the  first measurement of the $\tau$
lepton polarization in the decay $\bar B \to D^\ast \tau^- \bar \nu$ as well as a new measurement of
 in the hadronic
$\tau$
decay modes
which is statistically independent of the previous Belle
measurements,  with  a  different  background  composition \cite{Abdesselam:2016xqt}.
The  preliminary results
give \cite{Abdesselam:2016xqt}
\begin{equation}
R_{D^\ast} = 0.276\pm 0.034^{+0.029}_{-0.026}
\end{equation}
where the first errors are statistical and the second ones systematic.
This result
are  consistent with the theoretical predictions of the SM in Ref. \cite{Fajfer:2012vx}
 within 0.6$\sigma$ standard deviations.

At Belle II a better understanding of
backgrounds tails under the signal  and a reduction of the uncertainty  to 3\% for $R_{D^\ast}$  and 5\% for   $R_D$ is expected at 5 ab$^{-1}$.

While $R_B$ is defined as the ratio of branching fractions of decays that occur  at tree level in the SM at the lowest perturbative order, the observable  $R_K$ is defined as the ratio of branching fractions of rare decays. At LHCb,  $R_K$ has been measured to be \cite{Aaij:2014ora}
\begin{equation}
R_K \equiv  \frac{{\cal{B}}( B^+ \to K^+ \mu^+ \mu^-)}{{\cal{B}}(B^+ \to K^+ e^+ e^-)} =0.745^{+0.090}_{-0.074}\pm 0.036
\label{ratiotau0}
\end{equation}
where the first error is statistical and the second one is systematic. The measurement was performed
across the dilepton
invariant-mass-squared range 1 GeV$^2<m_{ll}^2 < 6$ GeV$^2$.
This result is 2.6$\sigma$ deviations away from the SM prediction $R_K^{SM}=1.0003 \pm 0.0001$
\cite{Bobeth:2007dw}. The impact of radiative corrections  has been estimated not to exceed a few \% \cite{Bordone:2016gaq}.

The alleged   breaking of lepton-flavour universality suggested by most of the data is quite large,  and several theoretical models have been tested against the experimental results.
A welcome feature of measurements in the $\tau$ sector is the capacity of putting  stringent limits on new physics models (see e.g. \cite{Celis:2012dk, Faroughy:2016osc, Becirevic:2016hea,Becirevic:2016yqi,Bordone:2016tex, Crivellin:2016ejn}).

\section{$|V_{cb}|$ and $|V_{ub}|$ determinations}

The   inclusive and exclusive semi-leptonic  searches  rely on
different theoretical calculations  and on
different experimental techniques which have, to a large extent, uncorrelated
statistical and systematic uncertainties. This independence makes
the comparison of $|V_{cb}|$ and $|V_{ub}|$ values from inclusive and exclusive decays an interesting test of our physical understanding (see e.g. \cite{Ricciardi:2016pmh,Ricciardi:2014aya,Ricciardi:2013xaa, Ricciardi:2014iga} and references therein).
Recent results of inclusive and exclusive determinations of $|V_{cb}|$ are collected in Table \ref{phidectab2}.
%
\begin{table}[h]
\centering
\caption{Status of exclusive  and  inclusive $|V_{cb}|$  determinations \cite{Ricciardi:2016pmh}}
\label{phidectab2}
\begin{tabular}{lrr}
\hline
 \hline
Exclusive decays &
( $ |V_{cb}| \times  10^{3}$) \\
\hline
 $\bar{B}\rightarrow D^\ast \, l \, \bar{\nu}$   & \\
\hline
FLAG 2016 \cite{Aoki:2016frl} & $ 39.27 \pm 0.49_{\mathrm{exp}} \pm 0.56_{\mathrm{latt}} $ \\
FNAL/MILC 2014 (Lattice $\omega=1$) \cite{Bailey:2014tva}   & $ 39.04 \pm 0.49_{\mathrm{exp}} \pm 0.53_{\mathrm{latt}} \pm 0.19_{\mathrm{QED}} $ \\
HFAG 2012 (Sum Rules) \cite{ Gambino:2010bp, Gambino:2012rd, Amhis:2012bh} & $   41.6\pm 0.6_{\mathrm{exp}}\pm 1.9_{\mathrm{th}} $ \\
\hline
 $  \bar{B}\rightarrow D \, l \, \bar{\nu} $  &   \\
\hline
Global fit   2016 \cite{Bigi:2016mdz}  &  $40.49 \pm 0.97$
\\
Belle 2015 (CLN)   \cite{Glattauer:2015teq,Lattice:2015rga}  & $ 39.86 \pm 1.33 $
 \\
Belle 2015 (BGL)   \cite{Glattauer:2015teq, Lattice:2015rga, Na:2015kha}  & $40.83 \pm 1.13$
 \\
FNAL/MILC  2015 (Lattice  $\omega \neq 1)$ \cite{Lattice:2015rga}  & $39.6 \pm 1.7_{\mathrm{exp+QCD}} \pm 0.2_{\mathrm{QED}} $\\
HPQCD  2015 (Lattice $\omega \neq 1)$  \cite{Na:2015kha}  & $40.2 \pm 1.7_{\mathrm{latt+stat}} \pm 1.3_{\mathrm{syst}}
$ \\
 \hline
Inclusive decays & \\
\hline
 Gambino et al. 2016 \cite{Gambino:2016jkc}  & $42.11 \pm 0.74$ \\
 HFAG  2014  \cite{Amhis:2014hma} & $ 42.46 \pm 0.88 $ \\
\hline
 Indirect fits & \\
\hline
UTfit  2016 \cite{Utfit16} &
$ 41.7 \pm  1.0 $
\\
CKMfitter  2015 ($3 \sigma$) \cite{CKMfitter} &
$41.80^{+0.97}_{-1.64}$
\\
\hline
\hline
\end{tabular}
\end{table}
The most precise estimates of $|V_{cb}|$ come from lattice determinations in the $B \to D^\ast$ channel, followed by determinations based on inclusive measurements. They all stay below 2\%  uncertainty.  We observe a tension between exclusive and inclusive determinations by comparing  the latest inclusive determination \cite{Gambino:2016jkc} and the latest  $B \to D^\ast$ FNAL/MILC  lattice result  \cite{Bailey:2014tva}, which amounts to about $3 \sigma$.
The tension lessens by comparing the same  inclusive determination with the (considerable less precise) exclusive   determination based on the sum rule calculation of the $B \to D^\ast$ form factor \cite{Gambino:2010bp, Gambino:2012rd, Amhis:2012bh}.
In the
$B \to D$ channel, where the uncertainty  has recently decreased, an inclusive/exclusive discrepancy is also observed  \cite{Bigi:2016mdz}.
It is  also possible to determine $|V_{cb}|$ indirectly, using
the CKM unitarity relations together with CP violation
and
flavor data, excluding direct informations on decays.
The indirect fits  provided  by the UTfit collaboration \cite{Utfit16} and the CKMfitter collaboration \cite{CKMfitter} are also reported in in Table \ref{phidectab2}.
Indirect fits  prefer a value for $|V_{cb}|$ that is closer to the (higher)
inclusive determination.

The parameter $|V_{ub}|$ is the less precisely known among the modules of the CKM matrix elements.
The CKM-suppressed decay $B \to \pi l \nu$ with light final leptons is the typical exclusive channel used to extract $|V_{ub}|$.
It is well-controlled experimentally and several measurements have been performed by both
 BaBar and Belle collaborations \cite{Hokuue:2006nr, Aubert:2006ry, Aubert:2008bf, delAmoSanchez:2010af, Ha:2010rf, Lees:2012vv, Sibidanov:2013rkk}.
Recently,
the measurements of branching ratios of
 $B \to \rho/\omega \, l \bar\nu_l$  decays and the computation of their form factors have been refined, and estimates of $|V_{ub}|$  have been inferred by these decays as well.
 Another channel depending on  $|V_{ub}|$ is the
baryonic  semileptonic $\Lambda^0_b \to p \mu^- \bar \nu_\mu$ decay. At the end of Run I,
LHCb has measured
 the probability
 of this decay relative to the channel $\Lambda^0_b \to \Lambda^+_c  \mu^- \bar \nu_\mu$ \cite{Aaij:2015bfa}.
This result has been combined with the ratio of form factors computed
using lattice QCD with
2+1
flavors of dynamical domain-wall fermions \cite{Detmold:2015aaa},
 enabling the first determination of the ratio of CKM elements $|V_{ub}|/|V_{cb}|$  from
baryonic decays \cite{Aaij:2015bfa}.
The value of $|V_{ub}|$ depends on the choice of the value of $|V_{cb}|$. By taking the inclusive determination $|V_{cb}|_{incl}= (42.21 \pm 0.78 )\times 10^{-3}$, the value
$|V_{ub}|= (3.50 \pm 0.17_{\mathrm{exp}} \pm 0.17_{\mathrm{FF}} \pm 0.06_{\mathrm{|V_{cb}|}} )\times 10^{-3}$ is obtained \cite{Rosner:2015wva}, where the errors are
from experiment, the form factors, and
$|V_{cb}|$, respectively.
By taking instead the higher value of the exclusive determination $|V_{cb}|= (39.5 \pm 0.8) \times 10^{-3}$, given by PDG 2014 \cite{Agashe:2014kda}, the LHCb reports \cite{Fiore:2015cmx}
$
|V_{ub}|= (3.27 \pm 0.23) \times 10^{-3}
$. The latter result, togheter with other recent exclusive determinations of $|V_{cb}|$, have been reported in Table
\ref{phidectab03}.
Let us observe that the values obtained for $B \to \rho/\omega \, l \nu$ appear to be systematically lower than the ones for  $B \to \pi l \nu$.
Indirect determination of $|V_{ub}|$   by the UTfit \cite{Utfit16}  and the CKMfitter \cite{CKMfitter} collaborations 
have also been reported in Table
\ref{phidectab03}.
At variance
with the $|V_{cb}|$ case, the results of the global fit prefer a value for $|V_{ub}|$ that is closer to the (lower)
exclusive  determination.
\begin{table}[h]
\centering
\caption{Status of  exclusive $|V_{ub}|$  determinations and indirect fits \cite{Ricciardi:2016pmh}}
\label{phidectab03}
\begin{tabular}{lrr}
\hline
Exclusive decays &  $ |V_{ub}| \times  10^{3}$
  \\
\hline
 $\bar B \rightarrow \pi l \bar \nu_l$      & \\
\hline
FLAG 2016    \cite{Aoki:2016frl}  & $3.62 \pm 0.14$\\
Fermilab/MILC  2015  \cite{Lattice:2015tia}  & $3.72 \pm 0.16$\\
RBC/UKQCD   2015  \cite{Flynn:2015mha}  & $3.61 \pm 0.32$\\
HFAG 2014 (lattice)    \cite{Amhis:2014hma}  & $3.28 \pm 0.29$\\
HFAG 2014 (LCSR)  \cite{Khodjamirian:2011ub, Amhis:2014hma}  & $3.53 \pm 0.29$   \\
Imsong et al. 2014 (LCSR, Bayes an.)     \cite{Imsong:2014oqa}  & $3.32^{+0.26}_{-0.22}$\\
Belle 2013 (lattice + LCSR)     \cite{Sibidanov:2013rkk}   & $3.52 \pm 0.29$ \\
\hline
 $\bar B \rightarrow \omega l \bar \nu_l$      & \\
\hline
Bharucha et al. 2015 (LCSR)     \cite{Straub:2015ica}  & $3.31 \pm 0.19_{\mathrm{exp}} \pm 0.30_{\mathrm{th}}$\\
\hline
 $\bar B \rightarrow \rho l \bar \nu_l$       & \\
\hline
Bharucha et al. 2015 (LCSR)     \cite{Straub:2015ica}   & $3.29 \pm 0.09_{\mathrm{exp}} \pm 0.20_{\mathrm{th}}$\\
\hline
 $ \Lambda_b \rightarrow p \, \mu\nu_\mu$      & \\
\hline
LHCb  (PDG)    \cite{Fiore:2015cmx}    & $ 3.27  \pm 0.23 $\\
\hline
Indirect fits &
\\
\hline
UTfit  (2016) \cite{Utfit16} &
$3.74 \pm  0.21$\\
CKMfitter  (2015, $3 \sigma$) \cite{CKMfitter} &
$ 3.71^{+0.17}_{-0.20}$
\\
\hline
\end{tabular}
\end{table}

The extraction of $|V_{ub}|$ from inclusive decays requires to address theoretical issues absent in the inclusive $|V_{cb}|$ determination, since the experimental  cuts, needed to reduce the background, enhance the relevance of the so-called threshold region in the phase space. Several theoretical schemes are available, which
are  tailored
to analyze data in the threshold region,  but  differ
in their treatment of perturbative corrections and the
parametrization of non-perturbative effects.
We limit to compare four theoretical different approaches, which have been recently analyzed
by BaBar \cite{Lees:2011fv}, Belle \cite{Urquijo:2009tp}  and  HFAG  \cite{Amhis:2014hma} collaborations, that is: ADFR by Aglietti, Di Lodovico, Ferrera and Ricciardi \cite{Aglietti:2004fz, Aglietti:2006yb,  Aglietti:2007ik}; BLNP
by Bosch, Lange, Neubert and Paz \cite{Lange:2005yw, Bosch:2004th, Bosch:2004cb}; DGE, the dressed gluon exponentiation, by Andersen and Gardi \cite{Andersen:2005mj}; GGOU by Gambino, Giordano, Ossola and Uraltsev \cite{Gambino:2007rp} \footnote{Recently, artificial neural networks have been used to parameterize the shape functions and  extract $|V_{ub}|$ in the GGOU framework \cite{Gambino:2016fdy}. The results are in good agreement with the original paper.}.
Although conceptually quite different, all these approaches
lead to roughly consistent results when the same inputs are used and the
theoretical errors are taken into account.
The HFAG estimates \cite{Amhis:2014hma}, together with the latest estimates by BaBar \cite{Lees:2011fv, Beleno:2013jla} and Belle
\cite{Urquijo:2009tp}, are reported in Table \ref{phidectab04}.
\begin{table}
\caption{Status of inclusive $|V_{ub}|$  determinations \cite{Ricciardi:2016pmh}}
\label{phidectab04}
\begin{tabular}{lrrrr}
 \hline
 Inclusive decays &
 $  |V_{ub}| \times  10^{3}$ &
  \\
\hline
& ADFR   \cite{Aglietti:2004fz, Aglietti:2006yb,  Aglietti:2007ik}  & BNLP    \cite{Lange:2005yw, Bosch:2004th, Bosch:2004cb}&  DGE    \cite{Andersen:2005mj} &   GGOU     \cite{Gambino:2007rp} \\
\hline
HFAG 2014 \cite{Amhis:2014hma} & $4.05 \pm 0.13^{+ 0.18}_{- 0.11}$ & $ 4.45 \pm 0.16^{+0.21}_{-0.22}  $  & $4.52 \pm 0.16^{+ 0.15}_{- 0.16}$ &
$4.51 \pm  0.16^{ + 0.12}_ { -0.15} $  \\
BaBar 2011  \cite{Lees:2011fv} &  $4.29 \pm 0.24^{+0.18}_{-0.19}  $  & $4.28 \pm 0.24^{+0.18}_{-0.20}  $    & $4.40 \pm 0.24^{+0.12}_{-0.13}  $
& $4.35 \pm 0.24^{+0.09}_{-0.10}  $ \\
 Belle 2009 \cite{Urquijo:2009tp} & $4.48 \pm 0.30^{+0.19}_{-0.19}  $ & $ 4.47 \pm 0.27^{+0.19}_{-0.21}  $ &  $4.60 \pm 0.27^{+0.11}_{-0.13}  $ & $4.54 \pm 0.27^{+0.10}_{-0.11}  $ \\
\hline
\end{tabular}
\end{table}
The BaBar and Belle  estimates  in Table \ref{phidectab04} refers to the value extracted by
the
most inclusive measurement, namely the one based on
the two-dimensional fit of the $M_X-q^2$
distribution with
no phase space restrictions, except for
$p^\ast_l > 1.0$  GeV. This selection  allow to access approximately
90\% of the total phase space \cite{Beleno:2013jla}.
The BaBar collaboration also
reports measurements of $|V_{ub}|$
in other regions of the phase space \cite{Lees:2011fv}, but the values reported in  Table \ref{phidectab04} are the most precise.
When averaged, the ADFR value is lower than the one obtained with the other three approaches, and closer to the exclusive values;  this difference
disappears
if we restrict to the BaBar
 and Belle results quoted in  Table \ref{phidectab04}.
By taking the arithmetic average of the
results obtained from these  four different QCD predictions of the partial rate the Babar collaboration gives \cite{Lees:2011fv}
$
|V_{ub}|=(4.33 \pm 0.24_{\mathrm{exp}} \pm 0.15_{\mathrm{th}}) \times 10^{-3}
\label{VinclBabar}
$.
%

By comparing the  results in Table \ref{phidectab03} and \ref{phidectab04}, we observe a tension between exclusive and inclusive determinations, of the order of $2-3\sigma$, according to the chosen values.
Belle II
is expected, at about 50 ab$^{-1}$,  to decrease experimental  errors on both inclusive and exclusive $|V_{ub}|$  determinations up to about 2\%.

\bibliography{VxbRef}

\end{document}